# Differences in Jazz Project Leaders' Competencies and Behaviors: A Preliminary Empirical Investigation

Sherlock A. Licorish and Stephen G. MacDonell
SERL, School of Computing & Mathematical Sciences
Auckland University of Technology
Private Bag 92006, Auckland 1142, New Zealand
sherlock.licorish@aut.ac.nz, stephen.macdonell@aut.ac.nz

## Abstract

*Studying the human factors that impact on software development, and assigning individuals with specific competencies and qualities to particular software roles, have been shown to aid software project performance. For instance, prior evidence suggests that extroverted software project leaders are most successful. Role assignment based on individuals' competencies and behaviors may be especially relevant in distributed software development contexts where teams are often affected by distance, cultural, and personality issues. Project leaders in these environments need to possess high levels of inter-personal, intra-personal and organizational competencies if they are to appropriately manage such issues and maintain positive project performance. With a view to understanding and explaining the specific competencies and behaviors that are required of project leaders in these settings, we used psycholinguistic and directed content analysis to study the way six successful IBM Rational Jazz leaders operated while coordinating their three distributed projects. Contrary to previous evidence reported in personality studies, our results did not reveal universal competencies and behaviors among these Jazz leaders. Instead, Jazz project leaders' competencies and behaviors varied with their project portfolio of tasks. Our findings suggest that a pragmatic approach that considers the nature of the software tasks being developed is likely to be a more effective strategy for assigning leaders to distributed software teams, as against a strategy that promotes a specific personality type. We discuss these findings and outline implications for distributed software project governance.*

## 1. Introduction

A growing body of evidence shows that the human factors involved during software development are some of the most influential determinants of project performance [1] and project outcomes. In particular, matching software practitioners to certain software roles has been shown to aid with task performance [2], the implication being that solving particular software tasks demands specific expertise, and individuals naturally showing higher levels of this expertise would perform better in corresponding roles. Such thinking has been extended to those involved in leadership positions, where it has been shown previously that successful leaders exhibited more extroversion than others [3-4]. Requirements such as this may be especially relevant in distributed software development contexts, where challenges related to distance, culture and personality demand that such teams are effective at self-organizing [5]. In this study we use data drawn from the IBM Rational Jazz repository to examine and explain the specific competencies and behaviors that are required of leaders of successful globally distributed projects when accounting for the nature of the software tasks they are governing. We provide insights into how these individuals hold their teams together, along with recommendations for the governance of other distributed teams.

In the next section (Section II) we present our theoretical background and related work, and outline our specific research question. We then provide our study design in Section III. In Section IV we present our results and analysis. Section V comprises a discussion of our findings, and in Section VI we consider our study's limitations. Finally, in Section VII we draw conclusions and outline our study's implications.

## 2. Theoretical Background and Related Work

Software human resource management has leveraged psychology and role theories in supporting the task of selecting individuals with appropriate skill sets for certain positions. Most software-related positions demand multiple capabilities, including intra-personal, organizational, inter-personal and management skills [6]. Intra-personal skills include judgment, innovation and creativity, and tenacity, while being self-organizing and having knowledge of specific environments (e.g., programming competence in Java) is characterized as organizational. Inter-personal skills comprise abilities in terms of teamwork, cooperation and negotiation, and management skills are related to planning, organization and leadership. In relation to software departments, roles may also be required due to the specific software process or methodology being utilized. For instance, a department adopting eXtreme Programming will likely necessitate roles such as programmer, tester, coach and so on [7]. Additionally, some environments may require that roles be performed by project members arbitrarily or as-needed, in which case team members must possess general competence in many roles [8]. Thus, while there may be some general conventions or assumptions regarding the suitability of individuals for particular software development tasks, role assignment and the competency needs of individual roles are at least in part subject to specific organizational requirements and contexts [9].

Apart from the consideration of competencies in the human resource management area, and the specific application of such competencies to software roles, software engineering as a discipline has also considered human involvement more generally in relation to development activities. The Software Engineering Institute provides the People-CMM to support the human dimensions of development [10], the personal software process (PSP) focuses on individual software practitioners' performance [11], and the team software process (TSP) [12] provides improvement guidance for software teams. Agile methodologies also substantially emphasize the importance of people elements in software development [13].

Psychology and role theories have also been applied with success in the software engineering and information systems disciplines [14]. Such considerations were founded on the view that human involvement, and the constraints that arise as a result of human issues in software development, substantially determine the outcomes of software projects. Thus, studying human elements of software development, such as software team formation and issues around how individuals behave in teams while solving problems, may provide valuable insights into critical success factors for software development. To this end, the study of software team formation and leadership has been approached from the standpoint of personality psychology [4], management psychology perception [15], and a combination of both perspectives [2].

For instance, Andre et al.'s Delphi study of role assignment and distribution in nine software teams [3] unearthed that successful teams were headed by leaders who favored planning as against a more fluid coordination approach, and such team leaders preferred a social and extroverted style when communicating. Wang [4] also found that project managers exhibiting extroversion tended to be successful. Similar findings have also been reported by Acuna et al. [2] based on their interviews with software project managers. However, this issue has received scant consideration for global software developments. Al-Ani et al. [16], one of the few works examining leadership in distributed teams, found team leaders' characteristics to be consistent across both co-located and distributed teams, tending to be more coordination-oriented.

Given recent growth in the adoption of this approach to software development [5], understanding the ways in which different leaders administer and coordinate their projects in these environments would inform distributed project governance. We suspect that, in these settings, the formal project leader may need to regularly compromise, perhaps encouraging, and releasing control to, informal leaders during specific project phases – and this may require specific competencies and behavioral orientations. Such findings would explain how distributed software leaders channel their project vision and explain their role(s) in their teams' self-organization. Ideally, observing those involved in the leadership of various types of software projects would help us to provide knowledge and recommendations for distributed team leadership across a range of project types [16]. Outcomes from such work would be especially useful if the leaders that are examined drive high-performing and successful teams, as these are conditions and outcomes others would seek to replicate. We therefore propose the following research question:

> *Are the competencies and behaviors of the leaders of globally distributed teams driven by the nature of their software tasks – or do successful leaders of distributed teams exhibit universal competencies and behaviors?*

## 3. STUDY DESIGN

During a case study to investigate the impact of Jazz (based on the IBM[R] Rational[R] Team Concert[TM] (RTC)[1]) project environments on team behaviors [17], we observed that a few individuals dominated project interactions (see Fig. 1 for a sample Jazz project communication network). Previous evidence has shown that those who interact the most are also the most important and influential in their teams' knowledge-sharing processes [18], and these individuals may occupy formal or informal leadership roles. We therefore followed previous work [19] and selected the top two communicators in each of three projects we had examined previously [17] (covering User Experience (UE), Coding (Code) and Project Management (PM) development tasks) to study project leaders' competencies and behaviors in cognizance of their project portfolio, using an embedded-case approach. First, we extracted the Jazz repository and mined the change histories to validate that our top communicators were indeed actual project leaders (see subsection A). We then used the messages contributed by these members to study their competencies and behaviors using linguistic analysis (see subsection B) and directed content analysis (see subsection C).

[1] IBM, the IBM logo, ibm.com, and Rational are trademarks or registered trademarks of International Business Machines Corporation in the United States, other countries, or both.

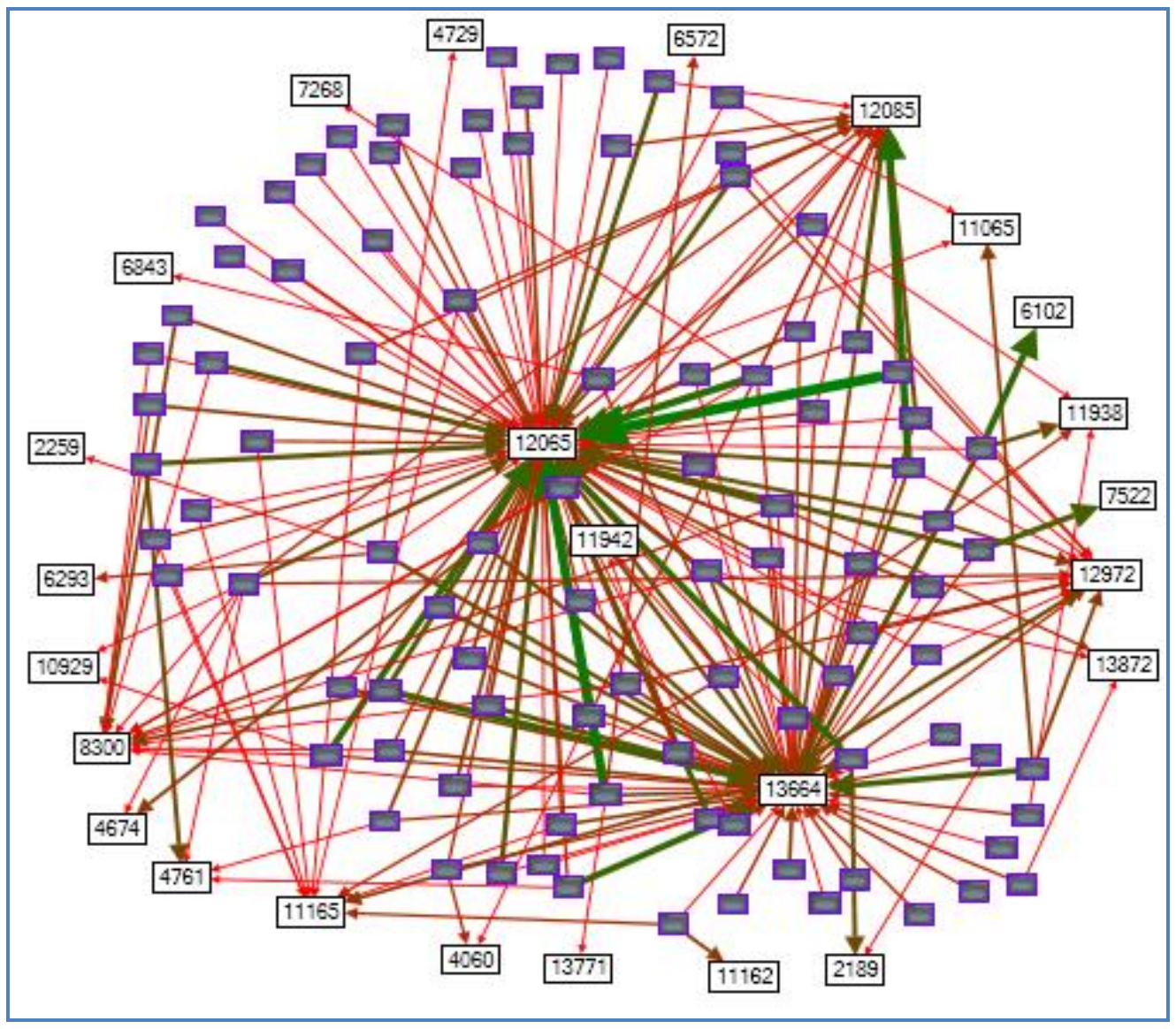

Figure 1. Directed network graph for a sample Jazz project showing highly dense network segments for practitioners "12065" and "13664"

### A. Case Repository and Data Collection

The repository examined in this work is a specific release of IBM Rational Jazz. Jazz, created by IBM, is a fully functional environment for developing software and managing the entire software development process [20]. The software includes features for work planning and traceability, software builds, code analysis, bug tracking and version control in one system. Changes to source code in the Jazz environment are only allowed as a consequence of a work item (WI) being created beforehand, such as a bug report, a new feature request or a request to enhance an existing feature. IBM afforded us an opportunity to study an instance (release 1.0.1) of the Jazz repository comprising a large amount of process data from development and management activities conducted across the USA, Canada and Europe. This release includes numerous projects, many of which are now commercially available and widely used (e.g., RTC). Jazz teams use the Eclipse-way methodology for guiding the software development process [20]. This methodology outlines iteration cycles that are six weeks in duration, comprising planning, development and stabilizing phases. Builds are executed after project iterations (also referred to as project milestones). All information for the software process is stored in a server repository, which is accessible through a web-based or Eclipse-based RTC client interface.

While the criteria for software project success are widely held to relate to their being completed on time, on budget and with the required features and functionality [21], others assert that measures related to a software project's impact on the developing organization, post-release customers' reviews and actual software usage are also relevant project success indicators [22]. Accordingly, given the impact that IBM Rational products (included in the Jazz repository) have had on IBM and many other organizations (with over 30,000 companies using these tools), and that these products have been positively reviewed and tested by those companies, we would infer that Jazz leaders and their projects should be considered as successful (see jazz.net for details). Thus, studying these team leaders should provide us with insights into successful leadership competencies and behaviors.

We previously created a Java program to leverage the Jazz Client API to extract team information along with development and communication artifacts from the Jazz repository (see Licorish and MacDonell [17] for details of the data extraction processes). For the three projects that were selected [17] (see Table I), we analyzed all the messages communicated by the six different project leaders (comprising a total of 708 messages altogether). Our sample of leaders, although small in size, is adequate for the type of inquiry we conducted, particularly as we selected all the study cases from a single homogeneous data source [23], and employed multiple analysis approaches (both top-down and contextual) in analyzing the large number of messages. Additionally, Romney et al. [24] show that samples comprising as few as four individuals can render highly accurate information if the individuals are very competent in the domain under investigation, as we contend is the case for Jazz leaders (noted earlier).

Of the total set of 708 messages, 258 were contributed by leaders working on UE tasks, 316 were posted by those leading Code work, and 134 were submitted by leaders working on PM components. Table II shows the project leaders' summarized change log activities, confirming that the members selected led their teams in task changes. Role information extracted from the repository revealed that four of the six members selected were actually assigned formal leadership roles while the remaining two were contributors (informal leaders).

TABLE I. SUMMARY STATISTICS FOR SELECTED PROJECTS

(UE - tasks related to developing UIs, Code - tasks associated with middleware development, PM - tasks were under the project managers' control)

| **Project ID** | **WI or Task Count** | **Project** | **Team Size** | **Total Messages** | **Period (days)** |
|---|---|---|---|---|---|
| UE | 54 | User Experience | 33 | 460 | 304 |
| Code | 207 | Code (Functionality) | 48 | 640 | 520 |
| PM | 210 | Project Management | 90 | 612 | 660 |
| ∑ | **471** | **-** | **171** | **1712** | **-** |

TABLE II. ACTIVITIES PERFORMED BY THE SELECTED PROJECT LEADERS ON SOFTWARE TASKS (PERCENTAGES)

| **Project ID** | **Tasks Created** | **Modifications Made** | **Tasks Resolved** |
|---|---|---|---|
| UE | 44.4 | 66.7 | 79.6 |
| Code | 39.6 | 85.0 | 93.7 |
| PM | 28.6 | 73.3 | 64.3 |
| **mean** | **37.5** | **75.0** | **79.2** |

### B. Linguistic Analysis Techniques

Previous research has identified that individual linguistic style is quite stable over time and that text analysis programs are able to accurately link language characteristics to competency and behavioral traits (see Mairesse et al. [25], for example). We therefore employed the Linguistic

Inquiry and Word Count (LIWC) software tool in our analysis of team competencies and behaviors. The LIWC was created after four decades of research using data collected across the USA, Canada and New Zealand [26]. Similar to an electronic parser, this tool accepts written text as input that is then processed based on the LIWC dictionary, after which summarized output is provided. The different linguistic dimensions considered by the tool and reported in the summary (see Table III) are said to capture the psychology of individuals by assessing the words they use [25-26]. For example, consider the following sample comment:

*"We are aiming to have all the patches ready by the end of this release; this will provide us some space for the next one. Also, we are extremely confident that similar bug-issues will not appear in the future."*

In this comment the author is expressing optimism that the team will succeed, and in the process finish ahead of time and with acceptable quality standards. In this quotation, the words "we" and "us" are indicators of team or collective focus, "all", "extremely" and "confident" are associated with certainty, while the words "some" and "appear" are indicators of tentative processes. Words such as "bug-issues" and "patches" are not included in the LIWC dictionary and would not affect the context of its use - whether it was to indicate a fault in software code or a problem with one's immunity to a disease. Although such omissions may be thought to represent a limitation of the LIWC, we *know* that the context is software development; and what is of interest, and is being captured by the tool, is evidence of competencies and behaviors. Previous work has provided confirmation of the utility of the LIWC dictionary for examining software developers' behaviors [17, 19]. However, apart from these studies, our search of the literature (covering ACM Digital Library, IEEE Xplore, EI Compendex, Inspec, ScienceDirect and Google Scholar) did not unearth any studies, in co-located settings or otherwise, employing the use of psycholinguistics. In the current work we examine team leaders' competencies and behaviors via their messages, along multiple linguistic categories. Table III depicts the categories chosen along with example terms and justification for their inclusion.

### C. Directed Content Analysis (CA)

We also studied the 708 messages contributed by the six leaders of the projects in Table I using a directed CA approach, with a hybrid classification scheme adapted from related prior work [31-32]. Use of a directed CA approach is appropriate when there is scope to extend or complement existing theories around a phenomenon [33], and so suited our further explorations of the leaders' competencies and behavioral processes. In fact, such examinations have led to enhanced understanding of many issues in the software engineering and information systems domains [34]. The directed content analyst approaches the data analysis process using existing theories to identify key concepts and definitions as initial coding categories. In our case, we used theories examining knowledge expressed during textual interaction [31-32] to inform our initial categories (Scales 1-8 in Table IV). Should existing theories prove insufficient to capture all relevant insights during preliminary coding, new categories and subcategories should be created [33]. Both authors of this work and two other trained coders categorized 5% of the communications (chosen at random) in a preliminary coding phase, and found that some aspects of Jazz leaders' utterances were not captured by the first version of our protocol (e.g., Instructions and Gratitude). During the pilot coding exercise we also found that these practitioners communicated multiple ideas in their messages. Thus, we segmented the communication using the sentence as the unit of analysis. We extended the protocol by deriving new scales directly from the pilot Jazz data (see Scales 9 to 13 in Table IV), after which the first author and the two trained coders recoded all the messages. Duplicate codes were assigned to utterances that demonstrated multiple forms of collaboration, and all coding differences were discussed and resolved by consensus (see Section IV.B for details). We noted an 81% inter-rater agreement between the three coders using Holsti's [35] coefficient of reliability measurement (C.R). This represents excellent agreement between the coders.

TABLE III. SELECTED LIWC LINGUISTIC CATEGORIES

| Linguistic Category | Abbrev. | Examples | Reason for Inclusion |
|---|---|---|---|
| Pronouns | I | I, me, mine, my, I'll, I've, myself, I'm | Elevated use of first person plural pronouns (we) is evident during shared situations, whereas, relatively high use of self references (I) has been linked to individualistic attitudes [27]. Use of the second person pronoun (you) may signal the degree to which members rely on (or delegate to) other team members [28]. |
| | we | we, us, our, we've, lets, we'd, we're, we'll | |
| | you | you, your, you'll, you've, y'all, you'd, yours, you're | |
| Cognitive language | insight | think, consider, determined, idea | Software teams were previously found to be most successful when many group members were highly cognitive and were natural solution providers [3]. These traits are also linked to effective task analysis and brainstorming capabilities. |
| | discrep | should, prefer, needed, regardless | |
| | tentat | maybe, perhaps, chance, hopeful | |
| | certain | definitely, always, extremely, certain | |
| Work and Achievement related language | work | feedback, goal, boss, overtime, program, delegate, duty, meeting | Individuals most concerned with task completion and achievement are said to reflect these traits during their communication. Such individuals are most concerned with task success, contributing and initiating ideas and knowledge towards task completion [29]. |
| | achieve | accomplish, attain, closure, resolve, obtain, finalize, fulfill, overcome, solve | |
| Leisure, social and positive language | leisure | club, movie, entertain, gym, party, jog, film | Individuals that are personal and social in nature are said to communicate positive emotion and social words and this trait is said to contribute towards an optimistic group climate [29]. Leisure related language may also be an indicator of a team-friendly atmosphere. |
| | social | give, buddy, love, explain, friend | |
| | posemo | beautiful, relax, perfect, glad, proud | |
| Negative language | negemo | afraid, bitch, hate, suck, dislike, shock, sorry, stupid, terrified | Negative emotion may affect team cohesiveness and group climate. This form of language shows discontent and resentment [30]. |

TABLE IV. CODING CATEGORIES FOR MEASURING INTERACTION

| Scale | Category | Characteristics and Example |
|---|---|---|
| 1 | Type I Question | Ask for information or requesting an answer – “Where should I start looking for the bug?” |
| 2 | Type II Question | Inquire, start a dialogue - “Shall we integrate the new feature into the current iteration, given the approaching deadline?” |
| 3 | Answer | Provide answer for information seeking questions - “The bug was noticed after integrating code change 305, you should start debugging here.” |
| 4 | Information sharing | Share information – “Just for your information, we successfully integrated change 305 last evening.” |
| 5 | Discussion | Elaborate, exchange, and express ideas or thoughts – “What is most intriguing in re-integrating this feature is how refactoring reveals issues even when no functional changes are made.” |
| 6 | Comment | Judgemental – “I disagree that refactoring may be considered the ultimate test of code quality.” |
| 7 | Reflection | Evaluation, self-appraisal of experience – “I found solving the problems in change 305 to be exhausting, but I learnt a few techniques that should be useful in the future.” |
| 8 | Scaffolding | Provide guidance and suggestions to others – “Let’s document the procedures that were involved in solving this problem 305, it may be quite useful.” |
| 9 | Instruction/ Command | Directive – “Solve task 234 in this iteration, there is quite a bit planned for the next.” |
| 10 | Gratitude/ Praise | Thankful or offering commendation – “Thanks for your suggestions, your advice actually worked.” |
| 11 | Off task | Communication not related to solving the task under consideration – “How was your weekend?” |
| 12 | Apology | Expressing regret or remorse – “Sorry for the oversight and the failure this has caused.” |
| 13 | Not Coded | Communication that does not fit codes 1 to 12. |

## 4. RESULTS AND ANALYSIS

### A. Linguistic Analysis

The three projects in Table I were selected following a purposive sampling approach, and thus, the results presented here should be considered in this context. We examined the data distributions for the three groups of leaders for normality using the Kolmogorov-Smirnov test and this showed violations of the normality assumption. We therefore conducted Kruskal-Wallis tests to check for differences in the 13 linguistic dimensions (shown in Table III) between leaders undertaking the three forms of software tasks (UE, Code and PM). These tests revealed that there were significant differences (P = 0.000) in language usage for each of the 13 linguistic dimensions (see the mean ranks and Kruskal-Wallis tests results shown in Table V) between leaders working on the three forms of tasks (UE, Code and PM). We therefore separated the data for the three groups of leaders and paired comparisons were done using Mann-Whitney U tests to reveal differences between type pairs (and these results are depicted in Table VI).

Table V shows that team leaders working on coding intensive tasks used the most individualistic (e.g., I, me, my) language; and Table VI shows that these differences were statistically significant (p = 0.000 and p = 0.000) when comparisons were made with those working on user experience and project management related tasks. These findings were similar for collective (e.g., we, our, us), social (e.g., give, buddy, love), negative (e.g., afraid, hate, dislike), cognitive (*insight*, *discrep*, *tentat* and *certain*) and leisure (e.g., club, movie, party) language use. Table V shows that reliance language (e.g., you, your, you’re) was rarely expressed by leaders working on the project management project; Table VI confirms that this was significantly different when compared to leaders’ use of reliance language on the other two projects (p = 0.000 and p = 0.000). Tables V and VI also show that leaders working on user experience tasks used significant amounts of positive language (e.g., beautiful, relax, perfect), and these individuals were the least focused on work (e.g., feedback, goal, delegate) and achievement (e.g., accomplish, attain, resolve).

TABLE V. MEAN RANKS AND KRUSKAL-WALLIS TEST RESULTS - LINGUISTIC PATTERNS OF LEADERS (PERCENTAGE USE)

| Linguistic Category | Abbrev. | Mean Rank | | | Kruskal-Wallis Test (*p*-value) |
|---|---|---|---|---|---|
| | | UE | Code | PM | |
| Pronouns | I | 279.85 | 453.31 | 265.22 | 0.000 |
| | we | 330.78 | 395.99 | 302.33 | 0.000 |
| | you | 370.13 | 362.02 | 306.67 | 0.000 |
| Cognitive | insight | 311.52 | 402.60 | 323.82 | 0.000 |
| | discrep | 298.10 | 424.77 | 297.38 | 0.000 |
| | tentat | 290.55 | 437.67 | 281.50 | 0.000 |
| | certain | 313.20 | 408.77 | 306.04 | 0.000 |
| Work and Achievement | work | 314.29 | 381.24 | 368.86 | 0.000 |
| | achieve | 288.18 | 405.14 | 362.76 | 0.000 |
| Leisure, social and positive | leisure | 330.56 | 380.18 | 340.05 | 0.000 |
| | social | 323.89 | 408.44 | 286.23 | 0.000 |
| | posemo | 442.54 | 300.68 | 311.91 | 0.000 |
| Negative | negemo | 324.99 | 393.34 | 319.71 | 0.000 |

### B. Directed CA

We coded the 708 messages belonging to the six leaders according to our protocol (shown in Table IV). In these messages we recorded 2191 utterances (UE = 648 codes, Code = 1245 codes, and PM = 298 codes). On average, leaders working on coding tasks communicated 3.9 different utterances in each message, those working on user experience tasks communicated 2.5 utterances, and those on the project management tasks said the least, with 2.2 utterances on average in each message. We present the aggregated codes in Table VII, which shows that Information sharing, Discussion, Scaffolding, Comments and Instructions dominated team leaders’ discourses. Given the differences in message counts (and codes obtained), we report the codes as percentages of the leaders’ overall utterances in each project in order to support comparisons (see Table VII).

In Table VII it is evident that there were differences among the leaders’ contributions in terms of Information sharing, Comments, Scaffolding, Instructions and Gratitude. The two leaders driving the project management related tasks tended to share Information relatively more frequently (58.4% of their communication compared to 42.1% and 48.5% for leaders working on user experience and coding tasks, respectively). Additionally, those working on coding related tasks used Comments proportionally more often (9.3% compared to 6.0% each for leaders working on user experience and project management tasks) along with Scaffolding utterances (15.5% compared to 7.9% and 6.4% for leaders working on user experience and project management tasks, respectively). Table VII shows that 11.7% of the utterances contributed by leaders working on user experience related tasks were Instructions, while 2.4% and 6.4% of this form of language were used by those working on coding and project management related tasks, respectively. A similar pattern is seen for Gratitude, which was only expressed by leaders working on user experience related tasks. A Chi-square test confirms that there were statistically significant differences in the way leaders working on the three projects contributed Information sharing, Comments, Scaffolding and Instructions to their projects' knowledge pools ($x^2 = 112.296$, df = 6, p = 0.000).

TABLE VI. MANN-WHITNEY TEST RESULTS - LINGUISTIC PATTERNS OF LEADERS (PERCENTAGE USE)

| Linguistic Category | Abbrev. | Mann-Whitney U Test (*p*-value) | | |
|---|---|---|---|---|
| | | UE - Code | UE - PM | Code - PM |
| Pronouns | I | 0.000 | 0.181 | 0.000 |
| | we | 0.000 | 0.052 | 0.000 |
| | you | 0.498 | 0.000 | 0.000 |
| Cognitive | insight | 0.000 | 0.604 | 0.000 |
| | discerp | 0.000 | 0.949 | 0.000 |
| | tentat | 0.000 | 0.463 | 0.000 |
| | certain | 0.000 | 0.405 | 0.000 |
| Work and Achievement | work | 0.000 | 0.011 | 0.751 |
| | achieve | 0.000 | 0.000 | 0.093 |
| Leisure, social and positive | leisure | 0.000 | 0.583 | 0.017 |
| | social | 0.000 | 0.088 | 0.000 |
| | posemo | 0.000 | 0.000 | 0.916 |
| Negative | negemo | 0.000 | 0.657 | 0.000 |

TABLE VII. CODING RESULTS FOR INTERACTION PATTERNS

| Category | ∑ Codes | | | % of utterances | | |
|---|---|---|---|---|---|---|
| | UE | Code | PM | UE | Code | PM |
| Type I Question | 20 | 14 | 3 | 3.1 | 1.1 | 1.0 |
| Type II Question | 37 | 50 | 11 | 5.7 | 4.0 | 3.7 |
| Answer | 31 | 44 | 19 | 4.8 | 3.5 | 6.4 |
| Information sharing | 273 | 604 | 174 | 42.1 | 48.5 | 58.4 |
| Discussion | 66 | 110 | 24 | 10.2 | 8.8 | 8.1 |
| Comment | 39 | 116 | 18 | 6.0 | 9.3 | 6.0 |
| Reflection | 16 | 39 | 5 | 2.5 | 3.1 | 1.7 |
| Scaffolding | 51 | 193 | 19 | 7.9 | 15.5 | 6.4 |
| Instruction/Command | 76 | 30 | 19 | 11.7 | 2.4 | 6.4 |
| Gratitude/Praise | 26 | 0 | 1 | 4.0 | 0.0 | 0.3 |
| Off task | 12 | 41 | 0 | 1.9 | 3.3 | 0.0 |
| Apology | 1 | 3 | 2 | 0.2 | 0.2 | 0.7 |
| Not Coded | 0 | 1 | 3 | 0.0 | 0.1 | 1.0 |
| **∑ Codes** | **648** | **1245** | **298** | - | - | - |

## C. Correlations – Linguistics and Directed CA

We ran a number of Pearson product-moment correlation tests to determine the relationship between team leaders' use of the various linguistic dimensions and the contribution of the CA categories over their projects (Spearman's Rank Order correlation tests produced similar results). We observe strong, positive correlations between the use of collective language (e.g., we, our, us) and the number of Questions asked, and our result was statistically significant ($p < 0.05$) for Type II Questions (Type I Questions: r = 0.508, n = 12, P = 0.092; Type II Questions: r = 0.669, n = 12, P = 0.017). We observe small and medium negative correlations between the use of reliance language (e.g., you, your, you're) and the number of Questions asked (Type I Questions: r = -0.161, n = 12, P = 0.618; Type II Questions: r = -0.330, n = 12, P = 0.295), and Answers provided (r = -0.180, n = 12, P = 0.576), although none of these relationships were significant. We found a strong, positive correlation between the amount of insightful language use (e.g., think, believe, consider) and leaders' sharing of Information, and this result was statistically significant (r = 0.708, n = 12, P = 0.010). We found a similar finding for insightful language and Scaffolding, although this particular result was not significant (r = 0.518, n = 12, P = 0.085). Results for insightful language use and Discussion and Comments were less strong, but also reflected positive correlations (r = 0.470, n = 12, P = 0.123; r = 0.345, n = 12, P = 0.272).

# 5. DISCUSSION

*Are the competencies and behaviors of the leaders of globally distributed teams driven by the nature of their software tasks – or do successful leaders of distributed teams exhibit universal competencies and behaviors?* While there appeared to be some similarities evident in the Jazz leaders' competencies and behaviors when governing specialized tasks, in general the project leaders' competencies and behaviors were divergent. Overall, we found that team leaders became more collective when they needed help, and there was greater group focus when these individuals were relying on others. We found that when there were higher levels of delegation among team leaders they engaged less with the wider team. We also observed that when leaders used more insightful language they offered more knowledge and guidance to their teams. These outcomes held across the three project types. We now examine the revealing competencies and behavioral differences among project leaders working on user experience, coding and project management tasks in the following three subsections, and discuss our findings in relation to the previous research introduced in Section II.

## A. User Experience

Jazz team leaders working on user experience tasks made extensive use of positive processes and were quite upbeat and optimistic, favoring a more extroverted outlook [30]. Comparatively speaking, these practitioners were not particularly achievement focused. Our results also show that leaders of the user experience project were most dependent on others. This evidence is corroborated with the lesser number of changes these members made to their team's cohort of project tasks, especially given the higher number of tasks these members created when compared to the other leaders (see Table II). The extensive use of positive processes may be linked to the nature of user experience tasks, such that leaders working in this team needed to be positive and encouraging when soliciting usability feedback from their team-mates. This type of positive language may be used in persuading the less active members to participate, a position supported by the higher level of gratitude expressed by these leaders, and the lower amount of work and achievement related processes these practitioners used. In fact, we observe that these leaders also instructed their team-mates twice as much as the other leaders (see Table VII), evidence that is linked to the lower level of actual task involvement in Table II. Firm instructions were perhaps given after more soft inter-personal communication [6] was unsuccessful in encouraging wider team cooperation. However, although these instructions were numerous (see Table VII), these leaders did not express high levels of frustration or anger when compared to those working on the coding tasks. The high levels of extroversion and positive behaviors observed among leaders working on user experience tasks are in line with previous observations [4].

### B. Coding

Team leaders working on coding tasks expressed the most pronounced competencies and behaviors of those considered, and these members appeared most conscientious (see Tables V and VII). Conscientiousness is linked to a general preference for order and goal-directed work [30] and is associated with success-driven individuals. Leaders working on coding intensive tasks were individualistic or self-focused, but these individuals also maintained high levels of group focus at times. The higher level of individualistic attitudes expressed by these members corresponds to the high level of task changes these members made (see Table II). However, although the coding task leaders handled a significant number of their team's features, they also maintained a collective or social team climate [29]. For instance, in addition to the high level of group focus, these members were most social, engaged the most about leisure when compared to the other leaders, and were the most verbose in terms of message content.

Similarly, leaders of the coding project used significant amounts of cognitive and negative processes, at much higher levels than the leaders of the other tasks. The computational nature of coding tasks demands substantial individual focus, and these Jazz leaders were also happy to lead from the front (as evident in the number of task changes they made). Task complexity has previously been found to impact individual performance [36]; the level of cognitive language used by leaders working on coding tasks, coupled with the higher amounts of debates and suggestions provided by these members (see Table VII) endorse the view that coding tasks were more demanding, and so required more organizational focus [6] and conscientiousness from the coding team's leaders. Finally, we observed that these leaders provided significantly more guidance and suggestion to their team through Scaffolding (see Characteristics and Example for Scale 8 in Table IV, and the results in Table VII). This in turn resulted in little need for them to provide their team with answers and instructions (as seen in Table VII). Coordination theories have shown that task programming (i.e., setting up mechanisms to automatically address repetitive tasks through tools, documentation and specifications) reduces the need for team engagement [22], and this was evident among the coding task leaders here.

### C. Project Management

Project leaders working on the project management tasks delegated the most, and also tended to talk comparatively more frequently about leisure. The relatively higher incidence of references to leisure denotes positive group climate and team spirit, necessary for motivating others [29]. However, these members were also work and achievement focused, and they shared substantially more information during their exchanges than the leaders working on the user experience and coding tasks. This finding is revealing as project management leaders were the least involved in actual task changes (see Table II), thus, there should be less need for them to provide their teams with context awareness (explanations and information about their cohort of features) and answers when compared to the others, and particularly, those involved in coding tasks (see Table VII). In fact, the much higher proportion of instructions provided by leaders on the project management tasks supports this position. Jazz leaders working on project management tasks seemed to express most compassion to their wider team-mates (although they said the least in each exchange), but they also exhibited organizational [6] and conscientiousness [30] traits, and offered instructions when this was needed. These members were also insightful (as seen in Table V), and while these tasks might be perceived to be less intensive than those related to coding, our evidence shows that leaders working on these tasks also demonstrated inter-personal competencies.

## 6. LIMITATIONS

*Measurement Validity*: The LIWC language constructs used to measure team behavior have been used previously to study this subject, and were assessed for validity and reliability [25]. However, the adequacy of these constructs in the specific context of software development warrants further investigation. Additionally, our contextual directed content analysis involving interpretation of textual data is subjective, and so questions naturally arise regarding the validity and reliability of the study outcomes. We employed multiple techniques to deal with these issues. First, our CA protocol was adapted from those previously employed and tested in the study of interaction and knowledge sharing [31-32], so there is a strong theoretical basis for its use. Second, we piloted the protocol and extended our instrument by deriving additional codes directly from the Jazz data. Third, we tested this extended protocol for accuracy, precision and objectivity, receiving an inter-rater measure indicative of excellent agreement [33]. Finally, our analyses did not consider the background of the practitioners (e.g., period of employment in leadership roles), which may influence the competencies and behaviors these members displayed.

*Internal and External Validity*: The tasks, history logs and messages from the three projects may not necessarily represent all IBM Rational Jazz leaders' processes in the repository. However, our evidence suggests that the practitioners selected are highly competent, and so, we believe that these members are likely to be reflective of those with similar orientations at IBM Rational Jazz [24]. Additionally, work processes and work culture at IBM are specific to that organization and may not be representative of organization dynamics elsewhere.

## 7. Conclusions and Implications

### A. Conclusions

Our findings demonstrate that practitioners developing software often operate within deep social and cognitive systems. Given these findings we believe that, in order for distributed software teams to succeed and navigate their projects' landscapes effectively, highly skilled team leaders are essential. The practitioners studied here demonstrated such skills. Jazz team leaders exhibited high levels of the social and inter-personal competencies that are essential to team work, cooperation and negotiation. However, competencies associated with judgment, innovation, creativity and self-organization were also deeply entrenched in these members' behaviors, and appeared more pronounced in some environments than others. Our next step is to extend this work using a larger sample of Jazz (and other) projects.

### B. Implications

Previous evidence shows that project leaders favoring a more extroverted and social outlook are most successful [3]. However, our findings suggest that a more pragmatic approach that considers the nature of the software tasks being developed and the level of team leaders' involvement is likely to be a more effective strategy for assigning leaders to distributed software teams. We found that, given the nature of the software task and team leadership involvement, while some leaders may well succeed due to high levels of extroversion and interpersonal competencies (e.g., for user experience tasks), for other tasks (e.g., coding) higher levels of conscientiousness and organizational competencies seem more useful, while leaders working on project management tasks demonstrated conscientiousness but also had organizational and inter-personal strengths. These findings are insightful as these Jazz leaders were equally successful at releasing features under their task portfolios.

Practitioners working in a distributed development context may tolerate some amounts of negative emotion and anger from their leaders when they are working on coding-intensive tasks due to empathy given their own experiences and expectations of the rigor involved in such tasks. However, it may be unsettling if these behaviors were expressed by leaders solving and coordinating user experience related tasks, where there seems to be generally more gratitude and positive behaviors. Additionally, given that usability engineers are often required to conduct usability testing involving the wider team, both for assessing ease of use and ensuring that features match previously planned requirements, if those leading usability efforts expressed anger and negative language, feedback from the wider team members may not be forthcoming, which may jeopardize software quality (as a consequence of inadequate feature evaluation and testing). This in turn could result in further costs for additional effort related to later rounds of feature enhancements. These unforeseen reworks are also likely to negatively impact previous release schedules and task allocation plans, affecting overall team performance.

### Acknowledgments

We thank IBM for granting us access to the Jazz repository. Thanks also to the coders for their help. S. Licorish is supported by an AUT VC Doctoral Scholarship Award.